\documentclass[usenatbib,usegraphicx]{mn2e}
\usepackage{subfigure}
\usepackage{longtable}
\usepackage{graphicx}
\usepackage{graphics}
\usepackage{keyval}
\usepackage{trig}
\def\beq{\begin{equation}}
\def\eeq{\end{equation}}
\def\bey{\begin{eqnarray}}
\def\eey{\end{eqnarray}}

\def\msun{M_\odot}
\def\sun{\odot}
\def\lsim{\mathrel{\raise.3ex\hbox{$<$\kern-.75em\lower1ex\hbox{$\sim$}}}}
\def\gsim{\mathrel{\raise.3ex\hbox{$  $\kern-.75em\lower1ex\hbox{$\sim$}}}}

\def\kms{\, {\rm km \, s}^{-1} }

\def\keV{\, {\rm keV} }

\title{X-ray group and cluster mass profiles in MOND: Unexplained mass on
the group scale}
\author[G. W. Angus, B. Famaey, D.A. Buote]{G. W. Angus$^{1}$\thanks{email:
gwa2@st-andrews.ac.uk}, B. Famaey$^{2}$, D.A. Buote$^{3}$\\   \\
$^{1}$SUPA, School of Physics and Astronomy, University of St. Andrews, KY16 9SS Scotland\\
$^{2}$Institut d'Astronomie et d'Astrophysique, Universit\'e Libre  
de Bruxelles,  
CP 226, Bvd du Triomphe, B-1050, Bruxelles, Belgium\\
$^{3}$Department of Physics and Astronomy, University of California Irvine, 4129 Frederick Reines Hall, Irvine, CA 92697-4575, USA\\  
}

\begin{document}

\date{Accepted ... Received ... ; in original form ...}

\pagerange{\pageref{firstpage}--\pageref{lastpage}} \pubyear{2007}

\maketitle

\label{firstpage}
\begin{abstract}
Although very successful in explaining the observed conspiracy between the baryonic distribution and the gravitational field in spiral galaxies without resorting to dark matter (DM), the modified Newtonian dynamics (MOND) paradigm still requires DM in X-ray bright systems. Here, to get a handle on the distribution and importance of this DM, and thus on its possible form, we deconstruct the mass profiles of 26 X-ray emitting systems in MOND, with temperatures ranging from 0.5 to 9~keV. Initially we compute the MOND dynamical mass as a function of radius, then subtract the known gas mass along with a component of galaxies which includes the cD galaxy with $M/L_K=1$. Next we test the compatibility of the required DM with ordinary massive neutrinos at the experimental limit of detection ($m_{\nu}=2$~eV), with density given by the Tremaine-Gunn limit. Even by considering that the neutrino density stays constant and maximal within the central 100 or 150~kpc (which is the absolute upper limit of a possible neutrino contribution there), we show that these neutrinos can never account for the required DM within this region. The natural corollary of this finding is that, whereas clusters (T$\ga$~3~keV) might have most of their mass accounted for if ordinary neutrinos have a 2~eV mass, groups (T$\lsim$~2~keV) cannot be explained by a 2~eV neutrino contribution. This means that, for instance, cluster baryonic dark matter (CBDM, Milgrom 2007) or even sterile neutrinos would present a more satisfactory solution to the problem of missing mass in MOND X-ray emitting systems. 


\end{abstract}


\section{Introduction}
\protect\label{sec:intr}
Recently obtained data from the cosmic microwave background (e.g., Spergel et al. 2007), large scale structure and type Ia supernovae would have us believe we inhabit a Universe dominated by some hitherto experimentally undetected dark matter requiring an extension to the current standard model of particle physics (cold dark matter, CDM) and something called dark energy, which no one can claim to understand, and which drives the acceleration of the expansion of the Universe (e.g., Perlmutter et al. 1999; Carroll 2004; Chernin et al. 2007). However, observations appear to thwart the extrapolation of this $\Lambda$CDM theory from its cosmological basis to the galaxy scale because they contradict an ever growing list of CDM predictions (e.g., Moore et al. 1999; Gentile et al. 2004; Kuzio de Naray et al. 2006; Bournaud et al. 2007). The latest gainsay of $\Lambda$CDM showed that not only can the predicted DM halos not be extrapolated down to galaxy scales (where myriad clauses in the unknown feedback of AGNs and supernovae provide freedom) but they conflict with massive clusters, where strong lensing shows the DM to be more concentrated than simulations predict (Broadhurst \& Barkana 2008). However, DM concentration-virial mass relations obtained from X-ray observations of relaxed groups and clusters agrees with $\Lambda$CDM simulations (Buote et al. 2007).

Actually, the observed conspiracy between the mass profiles of baryonic matter and dark matter at all radii in spiral galaxies (e.g. McGaugh et al. 2007; Famaey et al. 2007a) is more supportive of modified Newtonian dynamics (MOND, Milgrom 1983abc), a paradigm postulating that for accelerations below $a_o \approx 10^{-10} {\rm m} \, {\rm s}^{-2}$ the effective gravitational attraction approaches $(g_N a_o)^{1/2}$ where $g_N$ is the usual Newtonian gravitational field. 

Without resorting to CDM, this simple prescription is known to reproduce galaxy scaling relations such as the Tully-Fisher (e.g., McGaugh 2005) and Faber-Jackson (e.g., Sanders 2008), and the fundamental plane. More generally, it naturally explains the observed aforementioned conspiracy between the distribution of baryons and the gravitational field, in both low-surface brightness and high-surface brightness spiral galaxies. For reviews of serious work on MOND, see Sanders \& McGaugh (2002), Bekenstein (2006), and Milgrom (2008).

Recently, MOND has been pushed well beyond its original borders with the rotation curves of several tidal dwarf galaxies observed by Bournaud et al. (2007), requiring new molecular disk DM in Newtonian gravity but conforming strictly to the predictions of MOND (Gentile et al. 2007 and Milgrom 2007a). Furthermore, taking into account the possibility of varying the velocity dispersion anisotropy, MOND was shown to agree remarkably well with the line of sight velocity dispersions of the SDSS satellites (Angus et al. 2008), and the eight nearby MW dwarfs (Milgrom 1995; Angus 2008, but see the individual papers for caveats). Moreover, the theory successfullly predicted the local galactic escape speed from the solar neighbourhood (Famaey, Bruneton \& Zhao 2007b; Wu et al. 2007), and the first realistic simulations of galaxy merging were carried out (Nipoti et al. 2007; Tiret \& Combes 2007). 

Fundamental developments in the theory of gravity have also added plausibility to the case for modification of gravity through the work of, e.g., Bekenstein~(2004), Sanders~(2005), Zlosnik et al.~(2006, 2007a), Zhao~(2007), and Skordis~(2008), who have all presented Lorentz-covariant theories of gravity yielding a MOND behaviour in the appropriate limit by means of a dynamical vector field, which might arise from dimensional reduction of a higher-dimensional gravity theory (Bekenstein 2006; Mavromatos \& Sakellariadou 2007), or more generally from the fact that quantum gravity could define a preferred rest frame at the microscopic level (Jacobson 2008). Although rather fine-tuned and being somewhat of a long shot at producing a fundamental theory underpinning the MOND paradigm (see e.g. Bruneton \& Esposito-Farese 2007), these theories allow for new predictions, especially regarding cosmology. Surprisingly, such a vector field has precisely been shown to generate the instability that may produce large cosmic structures today in a MOND Universe (Dodelson \& Liguori 2006; Skordis et al. 2006;  Halle \& Zhao 2007; Zlosnik et al. 2007b). However, before trying to build a consistent cosmology (see Milgrom 2008 for possible alternative roots of MOND into cosmology), the MOND paradigm has yet to completely solve the mass discrepancy in X-ray emitting systems.

Groups of galaxies were studied by Milgrom (1998, 2002) by checking the stellar mass to light ratio required for consistency with the line of sight velocity dispersion of the group galaxies w.r.t. the centre of mass. This method found mass to light ratios of around a solar unit which bears the hallmark of no mass discrepancy, however, it could not probe the need for dark matter at smaller radii than these presumably largely separated galaxies. Of course, it is most probable that this absence of discrepancy is correct in the context of low mass groups with no detected X-ray emission, but in the case of X-ray bright groups and clusters a better gauge of dynamical mass is found using measurements of the properties of the X-ray gas (e.g., Gerbal et al. 1992; The \& White 1998; Sanders 1994, 1999, 2003, 2007; Aguirre et al. 2001;  Pointecouteau \& Silk 2005), or the combination of weak and strong lensing (e.g., Angus et al.\ 2007; Takahashi \& Chiba 2007; Famaey et al. 2007c; Milgrom \& Sanders 2008; Ferreras et al. 2008). The result is that MOND cannot as yet explain the mass discrepancy in X-ray emitting clusters of galaxies, and especially in their cores, while the gravitational lensing map of the bullet cluster has provided an extremely important constraint on the nature of all the missing mass, i.e. that it must be of a collisionless nature (Clowe et al. 2006; Bradac et al. 2006; Angus et al.\ 2007).

It has been conjectured (Sanders 2003, hereafter S03) that the mass discrepancy in galaxy clusters might be resolved by the addition of a component of massive neutrinos $m_{\nu}\sim$2eV, very near their maximum experimentally derived limiting mass of 2.2eV. These were indeed shown to be potentially consistent with the majority of clusters with temperature greater than 4~keV by S03, and with the bullet cluster (Angus et al. 2007). This hypothesis has the great advantage of naturally reproducing the proportionality of the electron density in the cores of clusters to $T^{3/2}$, as well as global scaling relations (Sanders 2007, hereafter S07). 
However, in a recent survey, Pointecouteau \& Silk (2005, hereafter PS05) studied a large sample of hot clusters ($>$4keV) and found that the central density of the dark matter was generally greater than allowed by the Tremaine-Gunn limit on neutrino density. This result was not damning, though, since the dynamical mass that could be accounted for by neutrinos under the Tremaine-Gunn limit was generally more than 90\%.


On the other hand, something that has never been addressed in the literature is the application of MOND to X-ray emitting groups and cool clusters in the range 0.5$<T<$3.0keV. The closest anyone has come is the study of the large elliptical NGC~720 by Buote \& Canizares (1994, 2002) which has T$\sim$0.6keV. They showed that the major axis of the isophotes of the X-ray emission were significantly offset from the major axis of the galaxy requiring dark matter of at least 4 times the mass of the visible matter. NGC~720 is currently the only elliptical galaxy for which this misalignment can be attributed to dark matter with reasonable confidence. A program to search for other very isolated, flattened elliptical galaxies that are sufficiently bright in X-rays for this type of study is underway.

Here we primarily use the high quality galaxy group data from Gastaldello et al. (2007), while the data for A2589 and NGC 4125 come from Zappacosta et al. (2006) and Humphrey et al. (2006) respectively. Our data for higher temperature clusters come from the published work of Vikhlinin et al. (2006). With these 26 systems, we perform a systematic check of the mass profiles of the MOND dark matter required for hydrostatic equilibrium of the X-ray emitting plasma, by subtracting the X-ray gas and the brightest cluster galaxy (BCG). We explore the correlation between the mass discrepancy in MOND and the masses and temperatures of the studied groups and clusters. We present the data in \S2 and derive the MOND dynamical and dark masses in \S3. In \S4, we substract the maximum contribution of conjectured 2eV neutrinos, by considering that the neutrino density stays constant, which is bound to {\it overestimate} the contribution that neutrinos might make. Even so, we confirm in \S5 that the central cores of hot clusters have densities that are too large to be consistent with ordinary 2eV neutrinos (PS05). More significantly and in support of the findings of Buote \& Canizares (1994, 2002), we show that all cool groups (0.5$<T<$3.0keV) have their MOND dynamical mass completely unaffected by these 2eV neutrinos, because groups are too cool to allow them to cluster densely enough even at the outskirts ($\sim 200$~kpc). In \S5.3, we discuss other possible solutions to solve the MOND mass discrepancy problem, notably cluster baryonic DM (CBDM, Milgrom 2007b, hereafter M07) and light sterile neutrinos in \S5.3 (e.g., Gentile, Zhao \& Famaey 2008).

\section{Data}
\protect\label{sec:data}

We use results obtained for relaxed clusters primarily from two recent studies of X-ray
clusters. These systems were chosen to be among the most relaxed known in order to ensure that hydrostatic equilibrium holds most accurately.

We use the published results of Vikhlinin et al. (2006) for
8 of the most massive clusters ($2 \, {\rm keV} < T < 9 \, {\rm
keV}$). For low-mass clusters (groups) we use results for 16 groups of
galaxies (1-2~keV) from the recent work of Gastaldello et al. (2007)
and another low-mass cluster (A2589) from Zappacosta et
al. (2006). Finally, we include the elliptical galaxy NGC 4125 ($T\sim
0.5$~keV) from Humphrey et al. (2006) to extend our sample of X-ray emitting systems down to the galaxy scale.

The objects in our sample are listed in Table 1. We refer the reader
to the references for details on the contruction of the density,
temperature, and Newtonian mass profiles. We note that for the massive
systems in the Vikhlinin et al. sample we exclude the central $\sim
20$~kpc from the analysis.  We do this for consistency with their
study, even though some of their systems (most notably A2029), do not
exhibit substantial morphological irregularities in their cores that
would lessen the validity of the approximation of hydrostatic
equilibrium.

\section{MOND dynamical mass in groups and clusters}
\protect\label{sec:mc}

To compute gravitating masses we assume the intracluster medium is
represented by a spherical single-phase ideal gas in hydrostatic
equilibrium. Although the X-ray isophotes of relaxed clusters are
approximately circular with modest ellipticity, the underlying mass
distribution is inferred to have substantial ellipiticy (0.4-0.6;
Buote \& Canizares 1996). Nevertheless, many previous studies have
shown that assuming spherical symmetry for relaxed clusters introduces
fairly small errors $\la 20\%$ which are acceptable for our purposes
(e.g., Tsai, Katz, \& Bertschinger 1994; Navarro et al. 1995; Buote \&
Canizares 1996; Evrard et al. 1996; Gavazzi 2005).

We derive the MOND dynamical mass for each system following the
approach of Sanders (1999). From the temperature and density profiles
of the hot gas, the centripetal gravitational acceleration, $g$, can
be deduced from the equation of hydrostatic equilibrium independently
of the gravitational theory:
\beq
\label{eqn:acc}
g(r)={-kT(r) \over {\rm w} m_p r}\left[ {d \ln \rho_X(r) \over d \ln r} + {d \ln T(r)\over d \ln r}  \right],
\eeq
where w is the mean molecular weight and w$m_p =5.2\times
10^{-58}\msun$ and the combination $kT(r)$ is in
units of keV. From this, the Newtonian dynamical mass can easily be
deduced. The studies cited in the previous section, from which we
obtain our data, use an NFW profile (Navarro, Frenk \& White 1995) to
match the dark matter profile.

Note that since we model clusters as spherical systems, we can ignore the
curl field of MOND (Bekenstein \& Milgrom 1984; Ciotti, Londrillo \&
Nipoti 2006; Angus, Famaey \& Zhao 2006), and since the external field
effect from Large Scale Structure (e.g., Angus \& McGaugh 2007; Famaey
et al. 2007b; Wu et al. 2007) is much smaller than the typical
gravitational acceleration in the region of interest, we can simply
use the relation of Milgrom (1983a) to derive the MOND dynamical mass:
\beq
M_m(r) = \frac{r^2}{G} g(r) \mu[g(r)/a_o],
\eeq
where $G=4.42\times10^{-3}{\rm pc}({\kms})^2 \msun^{-1}$ is
Newton's gravitational constant. Here the function $\mu (x)$ is chosen to be the simple function shown to have a good fit to the terminal velocity curve of the Milky Way
(Famaey \& Binney 2005) and to rotation curves of external galaxies
(Famaey et al. 2007a; Sanders \& Noordermeer 2007):
\beq
\label{eqn:simple}
\mu(x)=x/(1+x),
\eeq
while taking $\mu=1$ just reduces to the Newtonian dynamical mass.
This means that the MOND dynamical mass $M_m$ is related to the
Newtonian one $M_n$ by
\beq
\label{eqn:mondmass}
M_m (r) = \frac{M_n(r)}{1+a_o/g(r)},
\eeq
so that in MOND, there is a truncation of the dynamical mass
$M_m(r)$ at low accelerations. Here we take the MOND acceleration
constant to be $a_o=3.6 \, ({\rm km s^{-1}})^2 / {\rm pc}=1.2\times10^{-10}{\rm m}{\rm s}^{-2}$. Moreover, from Eq.~(\ref{eqn:mondmass}), one can see that at large radii, when $d\ln M_n/d\ln r < 1$ (as is the case for NFW profiles) and $g(r) << a_o$, then the MONDian enclosed dynamical mass begins to drop with radius. This is of course unphysical, and is caused by MOND generically predicting a logarithmic potential at large radii. This makes studies that do not consider the entire dynamical mass profile at all radii worthless when based on a Newtonian NFW profile. In this work, we circumvent this problem by flattening the enclosed MOND mass profile at its maximum value, and numbers in Table~1 are quoted at the radius corresponding to this maximum value. Further away, MOND predicts that the potential should not conform to an NFW profile but rather to an isothermal sphere.

For the objects in our sample from Gastaldello et al. (2007), Humphrey et al. (2006), 
and Zappacosta et al. (2006), we take the Newtonian gravitating mass
profiles used in those studies and simply extract the
MOND dynamical mass using Eq.~(\ref{eqn:mondmass}). Those studie
produced 20-30 Monte Carlo simulations of Newtonian mass profiles for
each object, which were used to produce the errors. For MOND profiles, because the $\mu$-function is not precisely known, any errors on the data are dwarved by the intrinsic errors. Therefore, we take the error on the MOND mass as half the difference between $M_m$ calculated with the simple $\mu$-function (Eq. \ref{eqn:simple}, which gives a relatively low mass) and $M_m$ with the standard $\mu$-function ($\mu=x/\sqrt{1+x^2}$, which gives a relatively high mass). These limits contain the majority of the reasonable MOND profiles and so can be naively taken as 1-$\sigma$.

Since we do not have error estimates for the density and temperature
profiles of the objects in the Vikhlinin et al. sample, we just use
the best-fitting values. However, as the objects are more massive and
generally brighter than the lower mass objects, we expect the relative
uncertainty on the massive systems to be typically comparable to or less than the lower mass
systems. In any case, the systematic errors on the MOND masses calculated by using different $\mu$-functions are large enough to account for any small random errors in the Newtonian mass. 

\subsection{Subtracting the X-ray gas and galaxies}

We integrated the plasma component to find the observed enclosed gas mass, $M_{X}(r)$ from
\beq
\label{eqn:mobs}
M_{X}(r)=\int_0^{r} 4\pi\rho_X(r)r^2dr
\eeq
where $\rho_X(r)$ is the density of plasma.


The empirical scheme for the galaxy mass profile employed by S03 and PS05, i.e. $M_*(r)/M_X(r) \approx 0.4(kT(r)/{\rm keV})^{-1}$ completely neglects the contribution of galaxies to the cluster mass below 150~kpc where the most massive galaxies reside. Furthermore, for groups this is a poorly motivated scheme and so for them we only include the mass of the BCG. This is reasonable as the MOND dynamical mass of groups saturates at around 100~kpc and we checked the surrounding 100~kpc of the BCG and found no galaxies contributing more than 5\% which is easily contained within the errors on $M/L_K$. 

For the clusters we used both the BCG luminosity and  the empirical galaxy density of S03 and PS05. The luminosity of the BCG in the K band from the Gastaldello et al. (2007) sample are taken from their paper, while those from Vikhlinin et al. (2006) are taken from Lin \& Mohr (2004) and the 2MASS survey. All are employed using a Hernquist profile. We subtracted this using $M/L_K$=1 which is standard for these very old stellar populations and a Kroupa IMF (Kroupa 2001); however, if a Salpeter IMF is used, we can expect $M/L_K$ to increase by around 50\% (Humphrey 2006). The X-ray studies of groups and clusters (Gastaldello et al. 2007; Zappacosta et al. 2006) typically find $M/L_K<$1, sometimes $<<1$ as they try to incorporate an NFW halo. By using $M/L_K$=1, we are making a conservative estimate of the amount of missing mass.

Subtracting both the X-ray mass and the galaxies from the total MOND dynamical mass leaves the total dark mass ($M_{DM}$) which is simply the mass of dark material necessary to reach agreement with the MOND dynamical mass.


\begin{figure*}
\def\subfigtopskip{0pt} 
\def\subfigbottomskip{4pt}
\def\subfigcapskip{1pt}
\centering

\begin{tabular}{cc}

\subfigure{\label{fig:n533}
\includegraphics[angle=0,width=8.0cm]{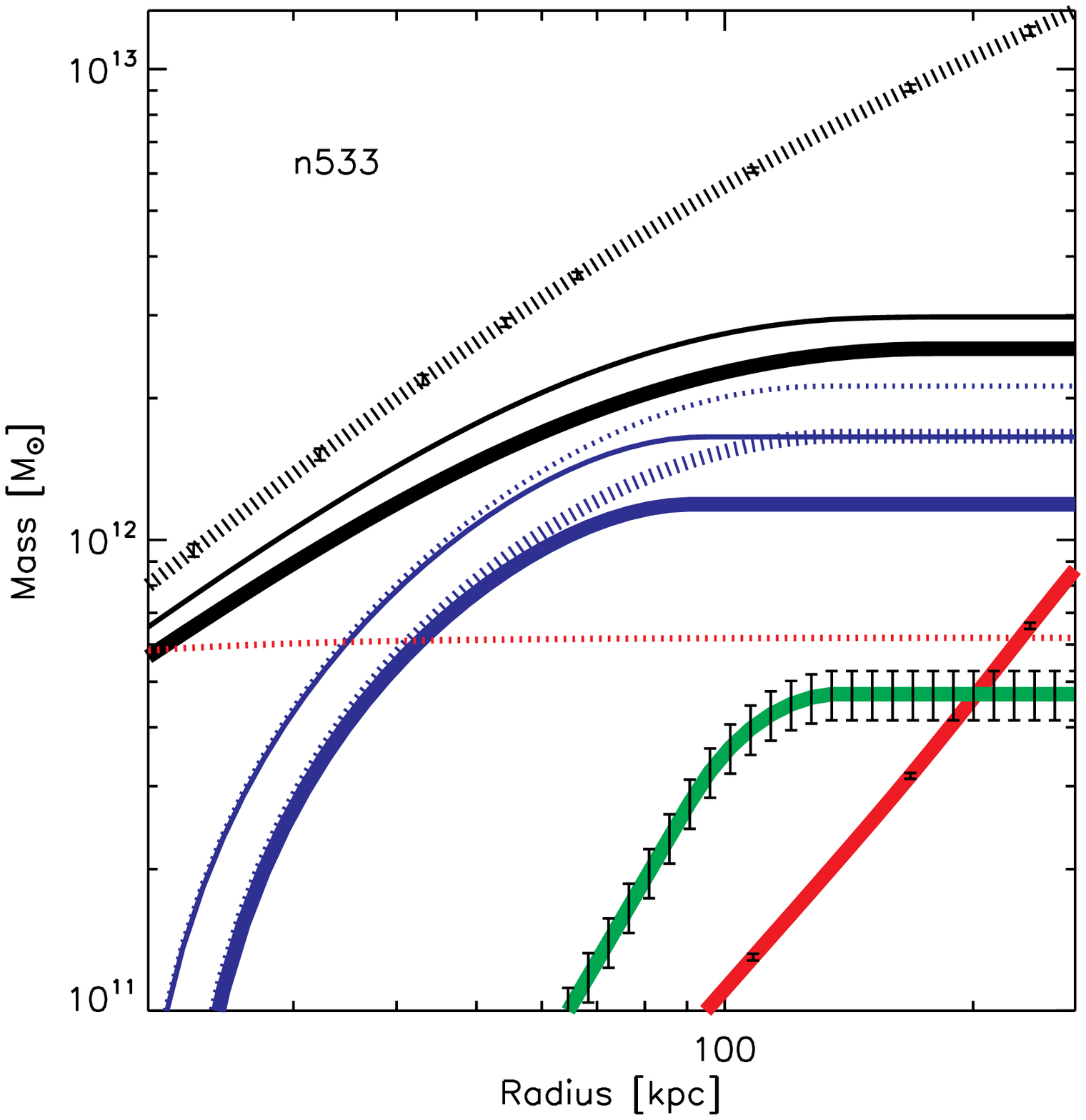}
}
&
\subfigure{\label{fig:n5044}
\includegraphics[angle=0,width=8.0cm]{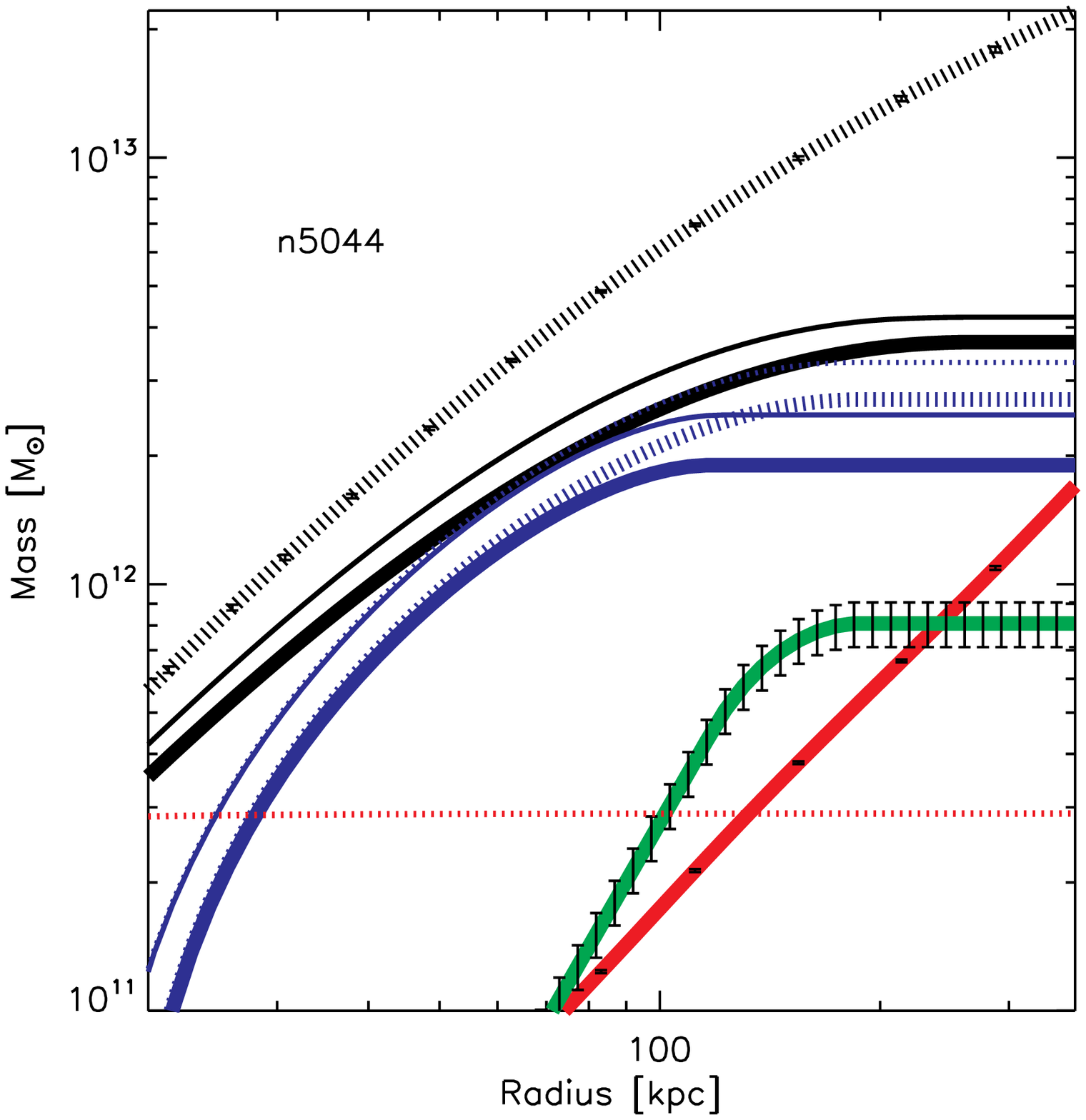}
}\\
\subfigure{\label{fig:a2717}
\includegraphics[angle=0,width=8.0cm]{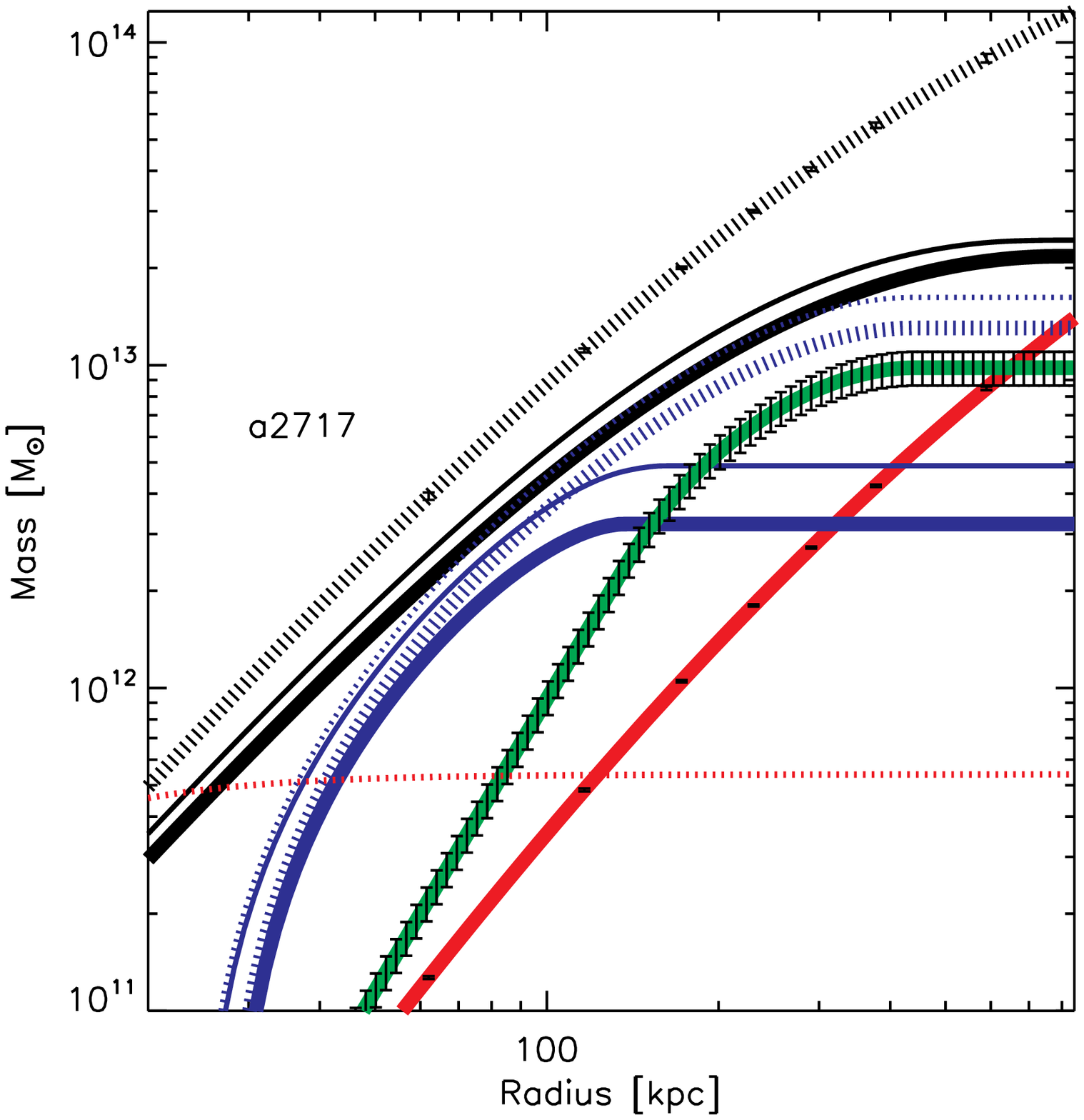}
}
&
\subfigure{\label{fig:a2029}
\includegraphics[angle=0,width=8.0cm]{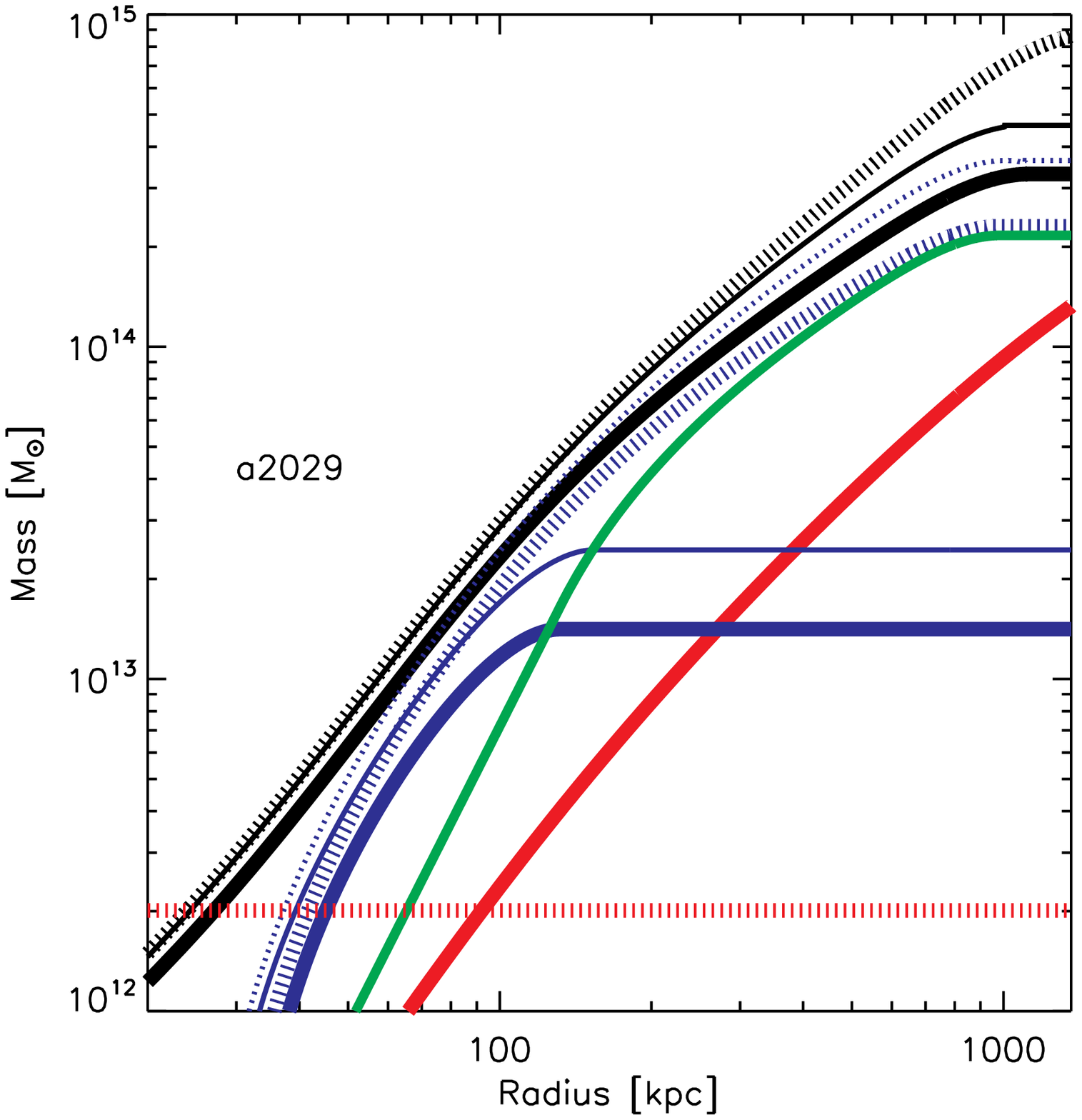}
}
\end{tabular}

\caption{Shows the mass profiles of the components of the dynamical mass for 4 representative clusters: N5044 (T=1.0~keV), N533 (T=1.2~keV), A2717 (T=2.2~keV),  and A2029 (T=8.5~keV). The total MOND dynamical mass ($M_m$) is in black with thick-solid linetype and its 1-$\sigma$ error is the thin-solid black line (there is virtually no error below). The Newtonian dynamical mass is black with a dotted linetype. The solid red line corresponds to the observed mass of X-rays and the dotted red line is the mass of the BCG if $M/L_K$=1 for which we use a Hernquist profile. The green line is the maximum necessary contribution of neutrinos; by this we mean it is maximal (Eq.\ref{eqn:tg}) in the centre where the dynamical mass is unexplained, whereas at the outskirts the neutrinos no longer have maximum density, instead they have the density necessary to complete the budget after gas and the residual mass at the centre has been accounted for. The thick solid blue line is the residual mass unexplained by the neutrinos and gas and its 1-$\sigma$ error is the thin solid blue line. The dotted blue line is the necessary residual mass if we have no significant neutrino density and again the 1-$\sigma$ error is the thin dotted blue line. Of course there is another error associated with the mass of the neutino, but we have fixed the neutrino mass at 2eV. Obviously a range of blue solid lines are possible between the solid blue and dotted blue depending on neutrino mass. If the neutrino has a mass $<<$2eV then we simply recover the dotted blue line. Clearly, one sees that for low temperature clusters such as N5044, no DM is present at r $>$ 150~kpc.  The errors on the different lines are not independent from each other. The firm  result here is that no random or systematic errors can ever make the green line (which overestimates the true contribution of 2~eV neutrinos by assuming a fully constant density in the core) reach the thick dotted blue line in the core of clusters, or even anywhere in groups, if the mass of ordinary neutrinos is smaller than its experimental upper bound.}
\label{fig:samp}
\end{figure*}

\section{Ordinary neutrinos as the MOND dark mass?}

At least two of the three active neutrinos ($\nu_{e}$, $\nu_{\tau}$ and $\nu_{\mu}$) have non-zero masses, meaning that they {\it must} be part of the mass budget of the Universe. Interestingly, in $\Lambda$CDM cosmology, it is possible to put
stringent limits on the sum of the three neutrino masses, from the angular power spectrum of the Cosmic Microwave Background and from the slope of the matter power spectrum (e.g., Zunckel \& Ferreira 2007; Host et al. 2007), essentially because a neutrino component that is too massive would not leave enough room for CDM in the matter budget of the Universe, resulting in the loss of small-scale power.

However, these constraints are not necessarily valid in a modified gravity cosmology (e.g., McGaugh 2004; Skordis et al. 2006) where the stronger gravity (covariantly in the form of an additional vector field) can effectively play the role of cosmic dark matter (thus not necessarily tracking the baryons, contrarily to the non-covariant case) by adding a new instability in the process of structure formation (e.g., Bekenstein 2004; Dodelson \& Liguori 2006; Halle \& Zhao 2007; Zlosnik et al. 2007b). 

Alternatively, model-independent experimental limits on the electron neutrino mass from the Mainz/Troitsk experiments, counting
the highest energy $\beta$-decay electrons of ${\rm ^3H} \rightarrow
{\rm ^3He}^{+} + {\rm e}^{-} + \nu_e + 18.57 \keV$ (the more massive
the neutrinos, the lower the cutoff energy of electrons), are $m_\nu<2.2eV$.  The KATRIN experiment (under construction) will be able to falsify 2eV neutrinos at 95\% confidence within months of taking data in 2009.

If the neutrino mass is substantially larger than the mass differences, then all types have about the same mass, and the cosmological density of three left-handed neutrinos and their antiparticles (e.g., S07) would be
\beq
\Omega_\nu = 0.062 m_\nu,
\eeq
where  $m_\nu$ is the mass of a single neutrino type in eV. If one assumes that clusters of galaxies respect the baryon-neutrino cosmological ratio, and that the MOND DM is mostly made of neutrinos as suggested by S03 and S07, then the mass of neutrinos must indeed be around 2~eV. 

In their modelling of the CMB anisotropies, Skordis et al. (2006) showed that such a component of 2eV neutrinos could actually prevent the MOND Universe from accelerating too much, and could thus yield the correct angular-distance relation to get the correct position of the peaks in the angular power spectrum of the CMB. 

The main limit on the neutrino ability to condense in clusters comes from the Tremaine-Gunn limit (Tremaine \& Gunn 1979), stating that the phase space density must be preserved during collapse. This is a density level half the quantum mechanical degeneracy level. The maximum density for a cluster of a given temperature, $T$, is defined for a given mass of one neutrino type as
\beq
\label{eqn:tg}
{ \rho_{\nu}^{max} \over  7\times 10^{-5} \msun pc^{-3}}=\left({T \over
1keV}\right)^{1.5}\left({m_{\nu} \over 2eV}\right)^4
\eeq
S03 showed that such 2~eV neutrinos at the limit of detection could indeed account for the bulk of the dynamical mass in his sample of galaxy clusters of $T > 4 \, {\rm keV}$ (see his Fig.8). Angus et al. (2007) also showed that such neutrinos could account for the weak lensing map of the bullet cluster (Clowe et al. 2006; Bradac et al. 2006), while S07 maintained that this hypothesis has the great advantage of naturally reproducing the proportionality of the electron density in the cores of clusters to $T^{3/2}$, as well as global scaling relations. However, looking at the central region of clusters, PS05 showed that neutrinos could not account for the dark matter all the way to the centre because the density reaches values much larger than the Tremaine-Gunn limit. However, this residual mass is only a few percent of the dynamical mass explainable by neutrinos.  

Here, we choose as a conservative approach to take the temperature of the neutrino fluid as being equal (due to violent relaxation) to the mean emission weighted temperature of the gas. We also assume that they contribute maximally, i.e. that their density is given by the Tremaine-Gunn limit, and constant (which is obviously untrue since the fluid obeys the equation of state of a partially degenerate neutrino gas, see e.g. S07). We thus {\it overestimate} the true contribution that neutrinos might make. This is of little concern, however, since, as we shall see in the next section, we do not wish to conclude anything beyond the fact that even overestimating the physical contribution of 2~eV neutrinos to the mass budget of the cluster does leave a massive, dense central component of unexplained mass, and contributes nothing to groups which again require a dense residual component. We shall estimate the importance of this residual component in clusters and groups, but beyond that we shall also consider the amount of DM needed if neutrinos have a negligible mass and do not contribute at all to the mass budget.

\section{Residual Mass}
\protect\label{sec:rm}
In Fig.~\ref{fig:samp} we plot for four representative objects several components of the group or cluster mass distinctly as functions of radius. The total MOND dynamical mass ($M_m$) is in black with solid linetype, the Newtonian dynamical mass ($M_n$) is black with a dotted linetype. The red line corresponds to the observed mass of X-rays ($M_X$) and the dotted red line is the mass of the BCG assuming $M/L_K$=1 and a Hernquist profile. The green line is the maximum necessary contribution of neutrinos ($M_\nu$). The solid blue line is what we call the residual mass ($M_{m-X-\nu-*}$) unexplained by the neutrinos, galaxies and gas, whereas the dotted blue line is the necessary total dark mass ($M_{m-X-*}=M_{DM}$) if we have no significant neutrino density. The various thin lines correspond to the systematic errors coming from the choice of $\mu$-function. The four representative clusters have respective temperatures 1.0, 1.2, 2.2 and 8.5~keV. Clearly, one sees that for low-temperature groups such as NGC5044, no DM is present at $r>150$~kpc.

We conclude that neutrinos might explain the mass discrepancy in the outer parts of the groups/clusters, especially those with high temperatures ($T>3$~keV), keeping in mind that no dark mass at all is necessary in the outer parts for the cooler clusters. However, to examine the real contribution of massive neutrinos in the outer parts of rich clusters, one should describe the equilibrium distribution expected for the neutrinos, and then examine whether this physically well-motivated profile can fit the data. This is however a non-trivial task in the context of MOND, needing to solve a Lame-Emden-like equation for polytropic models. Moreover, if the models are made more realistic by not considering purely self-gravitating neutrino spheres but also including a component of baryons, the task becomes even less trivial. This is far beyond the scope of the present paper, and will be the subject of further studies. 

\begin{figure}
\includegraphics[angle=0,width=8.5cm]{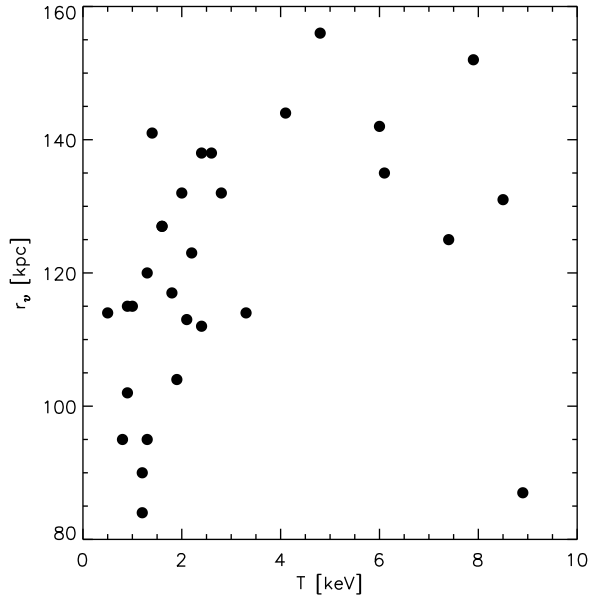}
\caption{Shows the radius below which neutrinos do not contribute in sufficient density to the MOND hidden mass.}\label{fig:rnut}
\end{figure}

Nevertheless, the important result is that even with our overestimate of the neutrino contribution, we found that neutrinos cannot account for the dark mass in the central 150~kpc. Much fuss has been made in the literature about the neutrinos being unable to account for this core of residual mass (PS05; Takahashi \& Chiba 2007, Ferreras et al. 2008), which is true. Indeed, every one of the sample clusters has a dominant central core of residual mass. Nevertheless, although all these systems have a residual dark component, it has been neglected as to how serious this extra component is, which the neutrinos cannot account for, and how this varies with temperature and mass of the cluster. Actually, in Fig. \ref{fig:mrmm}, which plots the ratio of residual mass (with and without neutrinos) to MOND dynamical mass for all clusters, we see that for the hot clusters studied by S03 and PS05 the residual mass (after neutrinos have been added) is a small fraction, with gas and neutrinos being more important globally (even though, as stated above, the contribution of neutrinos is a bit overestimated here). Conversely, for groups like NGC5044, the residual mass is completely dominant and is up to four times more significant than the stellar and gas components. Recall that the neutrinos cannot lay claim to the high densities in the cores due to the Tremaine-Gunn limit (and low temperatures of groups), but at a certain radius begin to dominate the dynamical mass. Given that our approximation is  bound to overestimate the neutrino contribution, this radius is actually the {\it minimal} radius at which neutrinos might begin to dominate the mass budget of the cluster. This radius $r_\nu$ is plotted as a function of temperature in Fig.\ref{fig:rnut} for 2eV neutrinos. It is intriguing that if we do live in a universe filled with $\sim$2eV neutrinos then the core sizes of residual DM in large and small clusters are very similar. Of course, if neutrinos are not present, then the residual DM is by far the most dominant component, moreso than the gas and galaxies for all groups and clusters. It is also important to realise that any departures from hydrostatic equilibrium generally cause underestimates of the total mass; i.e., from non-thermal pressure support. This would only exacerbate the problems highlighted in this paper.

\begin{figure}
\includegraphics[angle=0,width=8.5cm]{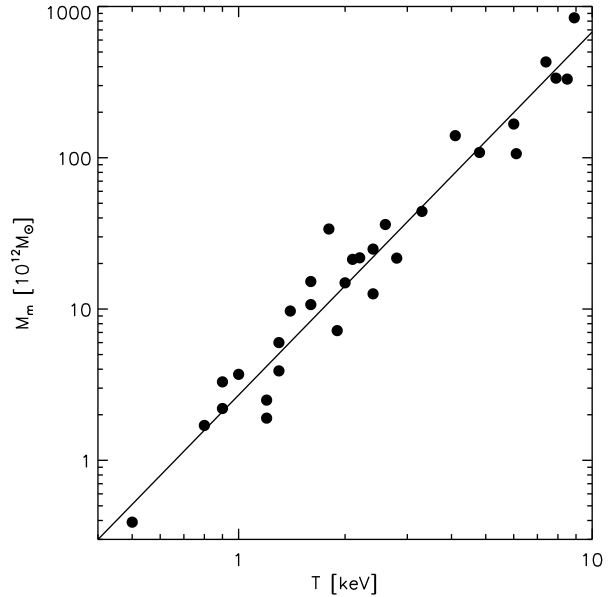}
\caption{Shows the scaling of the MOND dynamical mass with cluster temperature. The fitted line is $M_m \propto T^{2.4}$. This is higher than the MOND prediction of $M_m\propto T^2$ for the reasons explained in \S\ref{sec:rm}.}\label{fig:mmt}
\end{figure}

\begin{figure}
\includegraphics[angle=0,width=8.5cm]{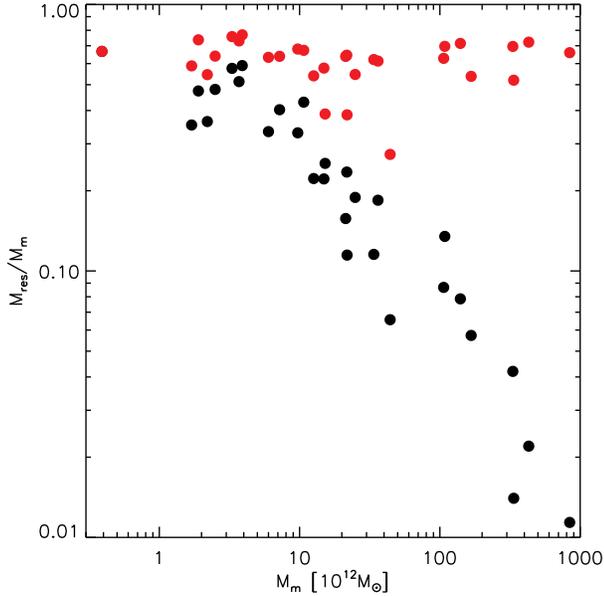}
\caption{Shows the fraction of dark mass (red) and of residual mass after neutrinos have been taken into account (black) as a function of the MOND dynamical mass. Clearly the ratio of dark mass stays constantly high ($>$0.6) if neutrinos are not contributing, but the fraction of residual mass becomes less important for heavier (and hotter) clusters when neutrinos are added.}\label{fig:mrmm}
\end{figure}

In Table 1 we list the different mass components in the
sample of clusters, and in Fig.\ref{fig:mmt} we plot the MOND dynamical mass of the clusters against temperature, which is fitted with a curve of the form $M_m \propto T^{2.4}$, close to the MOND prediction (S07) of a relation $M_m \propto T^{2}$. In order to get a strict $M_m \propto T^{2}$ the system has to reach the asymptotic MOND regime and the T used should be the asymptotic T, not the emission weighted average (or otherwise) and of course, must be constant. Furthermore, the systems should have the same logarithmic density slope at the measured radius.

\begin{figure}
\includegraphics[angle=0,width=8.5cm]{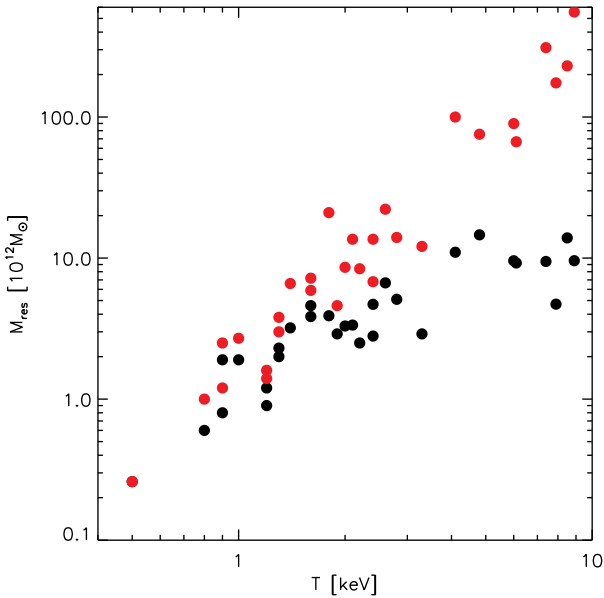}
\caption{Shows the residual mass discrepancy vs. temperature after subtraction of X-ray gas, the BCG and of the maximum contribution of neutrinos (black) and for the case with no neutrinos (red). In the case of significant neutrino contribution, this discrepancy is a fairly constant value of $\sim 10^{13}\msun$ for $T>3$~keV but drops steadily to zero for $T<3$~keV. If neutrinos are not present, the dark mass continues to rise with temperature.}\label{fig:mrt}
\end{figure}

We show in Fig.\ref{fig:mrmm} the fraction of residual mass as a function of cluster mass. If neutrinos are not present, then the fraction of residual dark mass is constantly high (between 60\% and 80\%) because it must account for the largest chunk of the dynamical mass, however, if neutrinos are present then the residual mass becomes less important for heavier clusters.  In Fig.\ref{fig:mrt}, one sees that the total dark mass scales with temperature, but that if 2~eV neutrinos are present the residual mass appears to saturate at a temperature of around 3~keV and a mass of $\sim 10^{13}\msun$. Interestingly, at the opposite end of the scale, the residual mass is steeply falling towards zero for groups of $T <2keV$, meaning that the amount of residual mass could in some sense be linked with the energy of X-rays themselves. Below the 0.5~keV threshold we get back to systems like the groups studied by Milgrom (1998, 2002), where no dark mass is needed and no X-rays are detected.

\subsection{Brightest Cluster Galaxies}

We express the quantity of residual mass in the central parts of the cluster as a MOND dynamical mass-to-light ratio for the BCG of the cluster, ranging between 1.7 and 20 (see Fig.\ref{fig:mlkt}). This mass-to-light ratio corresponds to an amount of mass that must be present within the radius $r_\nu$. This high dynamical mass-to-light ratio might perhaps be due to the interaction of this BCG with the neutrino fluid, allowing the fluid to heat and the neutrinos to clump more densely. From Fig.\ref{fig:mrlbcg} we see that the residual mass does indeed scale with the K-band luminosity of the BCG which is as expected if the neutrino fluid is heated allowing higher densities. If it was the case that the BCGs had large $M/L_K$ we should be able to fit the stellar mass profile (with variable normalisation) without a DM profile and it should describe the data well. Unfortunately, the mass profiles of galaxy groups/clusters are not well fit by a De Vaucoulers profile with $R_e$ set to that of the central galaxy but allowing for variable $M/L_K$. If we did this, we would over predict the interior mass profile. In addition, the barycentres of many clusters do not coincide with the BCG. This scaling of the residual mass with the luminosity of the BCG could thus rather mean that the heaviest BCGs reside in clusters with more X-ray emitting hot gas, and that large amounts of cold gas clouds are in turn also present (see \S~6.2).

\begin{figure}
\includegraphics[angle=0,width=8.5cm]{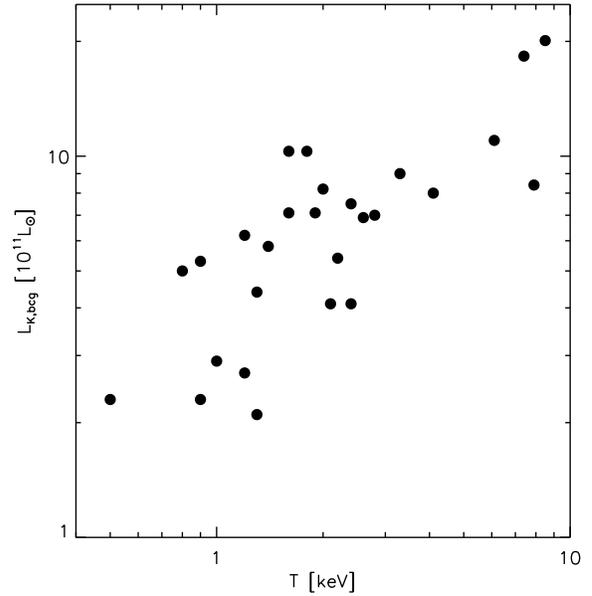}
\caption{Shows the marginal correlation of the K-band luminosity of the BCG with cluster temperature.}\label{fig:lbcgt}
\end{figure}

\begin{figure}
\includegraphics[angle=0,width=8.5cm]{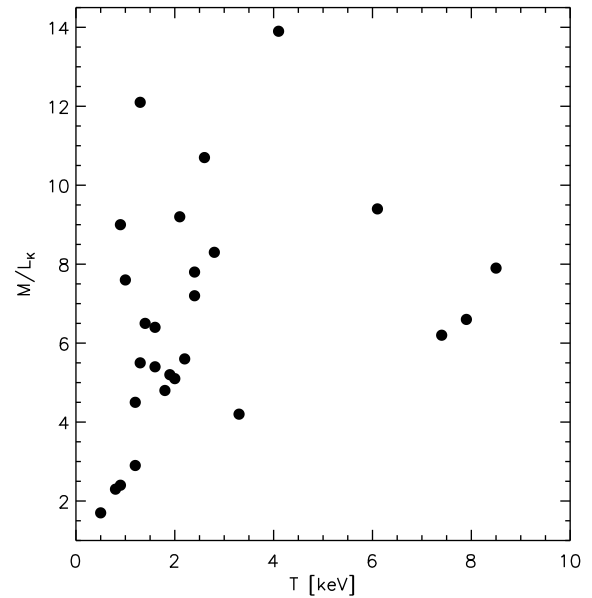}
\caption{Shows the necessary K-band M/L ratio of the BCG to explain the residual central mass discrepancy when neutrinos are present. It ranges between 2 and 20, meaning the stellar population of the BCG is probably not enough. If no neutrinos are present, then the ratio $>100$ for some clusters (see the table).}\label{fig:mlkt}
\end{figure}

\begin{figure}
\includegraphics[angle=0,width=8.5cm]{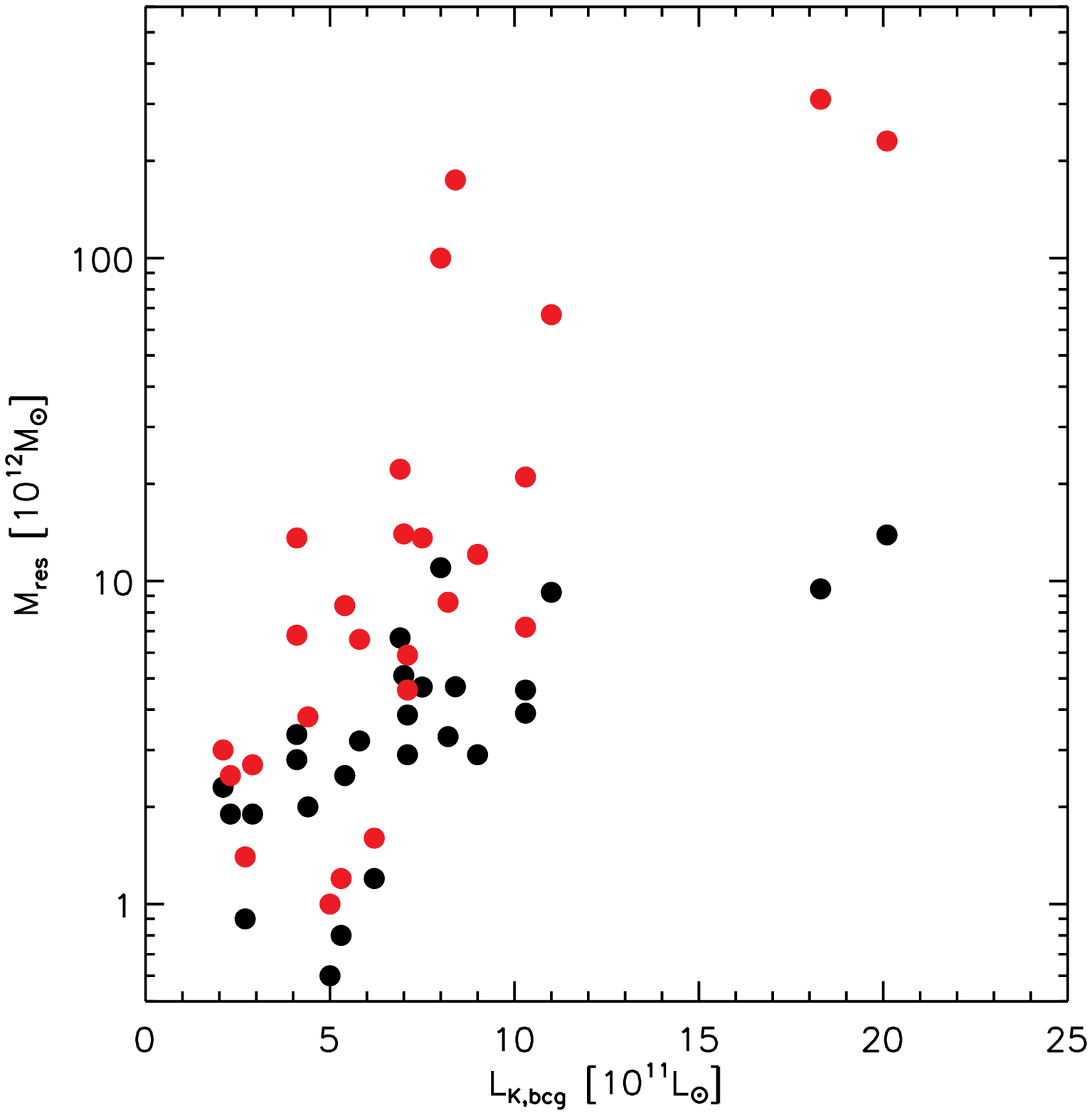}
\caption{Shows the scaling of the residual mass discrepancy as a function of BCG luminosity, with (black) and without (red) neutrinos. The scaling may reflect the deeper potential well heating the neutrino fluid allowing a higher density far beyond that allowed by the mean cluster temperature, or the fact that dark matter in the form of cold gas is more abundant in brighter galaxies. More trivially, it is just the natural expectation of brighter galaxies residing in more massive clusters.}\label{fig:mrlbcg}
\end{figure}

\subsection{Transition temperatures}
There is an increase in significance of the residual mass component left after the subtraction of gas and neutrinos for lower temperature clusters. Furthermore, for groups below 2keV, the contribution of neutrinos is negligible compared to the X-ray gas. Conversely, neutrinos have the potential to be of great importance in the dynamics of clusters hotter than 2keV, and subsequently are likely to help seed the collapse of these structures from cosmological perturbations (see Sanders 2008 for an introduction to galaxy formation in MOND).

It is well known that Romanowsky et al. (2003)'s stellar dynamical study of stellar tracers using planetary nebulae in purposely X-ray faint elliptical galaxies, generally do not require dark matter in the central parts (which is known as the problem of the dearth of dark matter in elliptical galaxies). Consequently, it is not surprising that they have also been shown to be consisent with MOND (Milgrom \& Sanders 2003, Tiret et al. 2007). In addition, the groups of galaxies studied by Milgrom (1998,2002) appear consistent with no dark matter. However, those studies suffer from
the well-known degeneracy from slight freedom in the M/L and velocity dispersion anisotropy, which
allows a wide range of mass profiles to be consisent with the data.

Angus et al. (2008) used the two cleverly assembled mock galaxy groups of Klypin \& Prada (2007) and computed the necessary M/L of the central galaxy in producing the los velocity dispersions of the stacked satellites. The lower mass host is around 8$\times10^{10}\msun$ using a sensible M/L in the g-band of 1.9-3.3, whereas the higher luminosity host requires 2.8$\times10^{11}\msun$ which can only be achieved with a M/L of 4.2-6.6 whereas the acceptable range is 3-5. So, there exists some evidence that the higher mass elliptical requires a moderate amount of DM.

We can only be certain that HSB galaxies like those studied by Sanders \& Noordermeer (2007) are fully consistent with MOND and no abundance of galactic DM. Below systems of this mass (at least) there is no need for DM in MOND. Studies of very X-ray luminous elliptical
galaxies have not generally examined the viablity of alternative
gravity theories. A recent exception is the study of the velocity dispersions of globular clusters within the giant elliptical NGC~1399 (Richtler et al. 2008), which has a two-temperature gas of 0.9-1.5~keV (Buote 2002), that has revealed some need for DM (within the aforementioned uncertainties on the anisotropy). Another exception is NGC~720, where the observed misalignment of the major axis of the stars and hot gas, along with the elongation of the X-ray isophotes, has been suggested to imply a
substantial mass discrepancy (Buote \& Canizares 1994; Buote et al. 2002). More recently the analysis of the large elliptical galaxy NGC~4125 (included here) with T=0.5keV shows more evidence for MOND DM in ellipticals. Here we find that we require NGC~4125 to have a minimum $M/L_K$=1.7 meaning, for the only system in our sample, there is less DM than luminous matter (note that this mass-to-light ratio in the K-band is obtained after subtracting the hot gas, and cannot be due to this component itself). 

The requirement of DM in MOND at 0.5~keV for NGC~4125, 0.6~keV for NGC~720 and the SDSS galaxies but no requirement for HSBs (Sanders \& Noordermeer 2007) or X-ray dim ellipticals and galaxy groups (Milgrom \& Sanders 2003, Milgrom 2002), necessitates a transition temperature below which the fraction of residual mass diminishes in importance and above which it increases in importance, peaks (between 0.5~keV and 2~keV) and then falls at higher temperatures if neutrinos are present (because the residual mass saturates at $10^{13}\msun$ but the dynamical mass continues to rise). If neutrinos do not contribute significantly, the residual amount of DM would directly scale with temperature. Why systems cooler than 0.5~keV show virtually no mass discrepancy in MOND must be related to the formation mechanism of such systems, and to the nature of the residual mass component found in X-ray groups (or of the whole MOND DM in groups and clusters if neutrinos have a negligible mass). Let us note that the very presence of DM in MOND seems to be synonymous with the presence of ionised gas and X-ray emission, which might hint at the presence of similar quantities of unseen cold gas that could perhaps explain this residual mass, and maybe even the whole of the mass discrepancy problem of MOND (M07).

\subsection{Cold gas clouds}
In the global baryon inventory of the Universe 50\% of the baryons produced during Big Bang nucleosynthesis (BBN) are still missing, and assumed to be in the warm-hot intergalactic medium (WHIM, see e.g. Roncarelli et al. 2006). Depending on the actual amount of baryons in the WHIM, there is thus some freedom for the residual mass to be baryonic in MOND. Indeed Fukugita \& Peebles~(2004) estimate that the observed baryons in clusters only account for 5\% of those produced during BBN, while it is clear from our Fig.~\ref{fig:samp} that the gas mass is increasing faster than the MOND dynamical mass, and tends to a DM/baryons ratio of about 1 at large radii. This means that if only 45\% of baryons are present in the WHIM, the remaining 5\% would be sufficient to solve the whole problem of missing mass in MOND clusters (even if neutrinos were found by KATRIN in 2009 to have mass much below the present-day experimental limit). One could thus say that, while CDM puts the missing baryons problem in individual galaxies, MOND has put it in galaxy clusters (see also McGaugh 2007). It should be highlighted again that DM in MOND only appears in systems with an abundance of ionised gas and X-ray emission (even in the case of galaxies such as NGC~4125). It is then no stretch of the imagination to surmise that these gas rich systems have equal quantities of molecular hydrogen (or other molecules), in e.g. some compact form to satisfy the collisionless citerion imposed by the bullet cluster.

M07 has recently proposed that the missing mass in MOND could entirely be ``cluster baryonic dark matter" (CBDM), e.g. in the form of cold, dense gas clouds. There is an extensive literature discussing searches for cold gas in the cores of galaxy clusters (e.g., Donahue 2006) but what is usually meant there is quite different from what is meant here, since those searches consisted in trying to find the signature of {\it diffuse} cold molecular gas at a temperature of $\sim 30$~K. The proposition of M07 rather relies on the work from Pfenniger \& Combes~(1994) and Walker \& Wardle~(1998), where dense gas clouds with a temperature of only a few Kelvins ($\sim3$~K), Solar System size, and of a Jupiter mass, were considered to be possible candidates for both galactic and extragalactic DM. Since the CBDM considered in the context of MOND cannot be present in galaxies, it is however not subject to the galactic constraints on such gas clouds. Note that the total sky cover factor of such clouds in the core of the clusters would be of the order of only $10^{-4}$, so that they would only occult a minor fraction of the X-rays emitted by the hot gas (and it would be a rather constant fraction). For the same reason, the chances of a given quasar having light absorbed by them is very small. Still, M07 notes that these clouds could be probed through X-ray flashes coming out of individual collisions between them.

Of course, this speculative idea also raises a number of questions, the most serious one being how these clumps form and stabilize, and why they form only in clusters, groups and some ellipticals, but not in individual spiral galaxies. As noted above, the fact that missing mass in MOND is necessarily associated with an abundance of ionised gas could be a hint at a formation and stabilization process somehow linked with the presence of hot gas and X-ray emission themselves. Then, there is the issue to know whether the clouds formation would be prior or posterior to the cluster formation. We note that a rather late formation mechanism could help increase the metal abundance, solving the problem of small-scale variations of metallicity in clusters when the clouds are destroyed (Morris \& Fabian 2003). M07 also noted that these clouds could alleviate the cooling flow conundrum, because whatever destroys them (e.g. cloud-cloud collisions and dynamical friction between the clouds and the hot gas) is conducive to heating the core gas, and thus preventing it from cooling too quickly. Such a heating source would not be transient and would be quite isotropic, contrary to AGN heating. This highly interesting and innovative scenario to alleviate the cooling flow problem should be the subject of further studies, but is however not the focus of our present paper.

\subsection{Sterile neutrinos}
In addition to the three flavours of ordinary neutrinos, there may be one or two sterile neutrinos of which the particle mass is neither constrained by the CMB or structure formation (e.g., Abazajian et al. 2001), and surely not in a modified gravity scenario. Such sterile neutrinos might be light enough (eV scale, see e.g. Maltoni \& Schwetz 2007) not to heavily perturb MOND fits in individual galaxies, but heavy enough to account for the missing mass in groups and clusters. Gentile et al.~(2008) postulated the existence of one or two species of 5~eV neutrinos in this context. Here, we computed the minimum sterile neutrino mass for a single species using the residual mass densities of NGC~533 and A2029 at 50~kpc, assuming that ordinary neutrinos have a negligible mass. We found that the minimal mass required was 5.6eV and 4.0eV respectively, meaning that the existence of a 6~eV sterile neutrino would have the potential to completely solve the cluster problem in MOND. However, further studies should describe the equilibrium distribution expected for such sterile neutrinos, and then examine whether physically well-motivated mass profiles can fit the data.

\subsection{Field oscillations}
Finally, we note that it has recently been suggested that non-trivial effects of the vector field in covariant formulations of MOND might create a DM-like effect (e.g., Zhao 2007), alleviating the need for neutrinos and/or CBDM. It was argued that this effect was most important at the tidal boundary of merging, rapidly varying, systems, such as the high collision speed of the bullet cluster. It however remains to be seen how this effect would offer an explanation to the MOND mass discrepancy found for the quasi-static systems studied in this paper, including X-ray emitting elliptical galaxies. Alternatively, the introduction of an additional scalar field (e.g., Sanders 2005) in covariant theories of MOND can induce the presence of bosonic DM (through the scalar field oscilations) that cannot accumulate in spiral galaxies, but that could help the cause of the missing mass in groups and clusters.

\section{Conclusion}

We have used high quality data to decompose the mass profiles of 26 X-ray emitting systems in MOND, with temperatures ranging from 0.5 to 9~keV. We confirmed that galaxy groups and clusters in MOND require a huge DM component to make up for all the dynamical mass (60-80\% of the total dynamical mass at the last observed radius, dropping thereafter). We have shown that, whatever their precise equilibrium distribution, hypothetical 2~eV ordinary neutrinos can never explain a component of residual mass enclosed within the central 100 or 150~kpc. This confirms the result of PS05, but we add an important corollary to this finding, namely that groups with T$\lsim$~2~keV cannot be explained by a 2~eV neutrino contribution. Indeed, we have probed far lower masses ($<10^{13}\msun$) than PS05, and have shown that this residual mass component, which is present in all known X-ray bright systems hotter than 0.5keV, is the dominant one for groups with T$<2keV$.

This leads us to believe that, if MOND is correct, the current baryon inventory of galaxy clusters is unfinished and that the unaccounted for baryons known to exist from arguments of Big Bang Nucleosynthesis (depending on the fraction residing in the WHIM) are indeed buried in the central regions of clusters in some hitherto undetectable form. Constraints from the bullet cluster tell us that this mass must be compact and cool. This could hint at dense cold gas clouds as recently suggested by M07. It could alternatively hint at the existence of a 4th light sterile neutrino of $\sim$~6eV mass.

\section{acknowledgements}
The authors thank Fabio Gastaldello for his low temperature cluster data, Philip Humphrey for the NGC4125 data and Luca Zappacosta for providing the A2589 data from his paper. We also thank Moti Milgrom, Francoise Combes, Maxime Robert and Luca Ciotti for comments on the manuscript. GWA acknowledges his PPARC scholarship and BF is an FNRS research associate.

\begin{appendix} 
\begin{table*} 
\centering
\begin{tabular}[p]{|c|c|c|c|c|c|c|c|c|c|c|c|c|c|}
${\rm Cluster }$ &${\rm T }$ & ${\rm r_{max}}$&$ {\rm M_{\Delta}}$&$ {\rm 
M_X}$ &${\rm M_m}$&${\rm M_{DM}}$&$ {\rm M}_\nu$ & ${\rm r}_\nu$ & $M_{m-X-\nu}$ & $L_{K,bcg}$ & $\Upsilon_{K,bcg}$&Ref\\ 
$ $ & $ {\rm keV} $ & ${\rm kpc} $& ${\rm 10^{13} \msun} $ & ${\rm 10^{12} \msun } $ & ${\rm 10^{12} \msun } $ &${\rm 10^{12} \msun } $& ${\rm 10^{12} \msun} $ &${\rm kpc } $&${\rm 10^{12} \msun } $ &${\rm 10^{11} L_{\sun} } $& \\ 
(1)& (2) & (3) & (4) & (5) & (6) & (7) & (8) & (9) & (10)&(11)&(12)&(13) \\ 
      NGC4125  &      0.5  &       460  &    $0.6\pm 0.1^c$ &      0.0  &      0.39 &       0.26&      0.0  &      114  &       0.16  &      2.3  &        1.7 (1.7)&(d)  \\ 
     NGC5129  &      0.8  &       278  &     $0.8\pm0.1^a$  &      0.8  &      1.7  &      1.0  &      0.3  &       95  &       0.6  &       5.0  &       2.3 (2.9) &(a)  \\
     RGH80  &      0.9  &       521  &     $1.9\pm0.1^d$  &      4.7  &      2.2  &      1.2  &      0.4  &      102  &       0.8  &       5.3  &       2.4 (3.2) &(a)  \\
     NGC4325  &      0.9  &       225  &     $1.3\pm0.2^b$  &      0.7  &      3.3  &      2.5  &      0.7  &      115  &       1.9  &       2.3  &       9.0 (11.9)&(a)   \\
     NGC5044  &      1.0  &       369  &     $1.9\pm0.1^a$  &      1.7  &      3.7  &      2.7  &      0.8  &      115  &       1.9  &       2.9  &       7.6 (10.4)&(a)   \\
     NGC2563  &      1.2  &       252  &     $0.9\pm0.1^b$  &      0.6  &      1.9  &      1.4  &      0.4  &       84  &       0.9  &       2.7  &       4.5 (6.1) &(a)  \\
      NGC533  &      1.2  &       265  &     $1.3\pm0.1^a$  &      0.9  &      2.5  &      1.6  &      0.5  &       90  &       1.2  &       6.2  &       2.9 (3.6)  &(a) \\
     NGC1550  &      1.3  &       216  &    $1.4\pm0.1^b$   &      1.0  &      3.9  &      3.0  &      0.6  &       95  &       2.3  &       2.1  &      12.1 (15.1) &(a)  \\
    MS0116  &      1.4  &       405  &     $4.9\pm1.6^a$  &      2.1  &      9.7  &      6.6  &      3.5  &      141  &       3.2  &       5.8  &       6.5 (12.4) &(a)  \\
   RXJ1159  &      1.6  &       708  &     $6.1\pm3.3^d$  &      7.2  &     10.7  &      7.2  &      2.7  &      127  &       4.6  &      10.3  &       5.4 (8.0)&(a)   \\
   RXJ1159  &      1.8  &       700  & $3.0\pm0.3^b$  &     15.6    &     33.8    &      21.0    &      17.1    &      117  &      3.9  &      10.3  &       4.8 (21.4) &(b)   \\
      MKW4  &      1.9  &       337  &     $3.2\pm0.1^a$  &      2.5  &      7.2  &      4.6  &      1.6  &      104  &       2.9  &       7.1  &       5.2 (7.4) &(a)  \\
      MKW4  &      1.6  &       634  &     $2.8\pm0.1^a$ &     19.3  &     15.2  &      5.9  &      2.0  &      127  &      3.8  &       7.1  &       6.4  (9.3) &(b)   \\     
    ESO552  &      2.0  &       484  &     $5.5\pm0.5^a$ &      4.0  &     14.9  &      8.6  &      5.3  &      132  &       3.3  &       8.2  &       5.1 (11.5)&(a)   \\
      A0262  &      2.1  &       650  &     $3.4\pm0.5^b$   &    14.4  &     21.3  &     13.6  &      10.2 &      113  &      3.3  &       4.1  &       9.2  (34.2) &(b)   \\
      A0262  &      2.4  &       276  &    $3.6\pm0.1^b$  &      2.4  &     12.6  &      6.8  &      4.0  &      112  &       2.8  &       4.1  &       7.8 (17.6) &(a)  \\
     A2717  &      2.2  &       840  &    $10.7\pm0.5^d$  &     14.0  &     21.8  &      8.4  &      5.9  &      123  &       2.5  &       5.4  &       5.6 (16.5) &(a)  \\
      AWM4  &      2.4  &       524  &     $7.4\pm0.6^a$  &      5.8  &     24.9  &     13.6  &      8.9  &      138  &       4.7  &       7.5  &       7.2 (19.1) &(a)  \\
     A1991  &      2.6  &       732  &$12.3\pm1.7^d$  &     59.8    &     36.2    &      22.2    &      15.5    &      138  &      6.6  &       6.9  &      10.7 (33.1) &(b)   \\
    ESO306  &      2.8  &       284  &     $6.0\pm1.1^b$  &      2.5  &     21.7  &     14.0  &      8.9  &      132  &       5.1  &       7.0  &       8.3 (21.0)  &(a) \\
     A2589  &      3.3  &       779  &   $33.0\pm 3.0^c$  &     18.3  &     44.2  &     12.1  &      9.2  &      114  &       2.9  &       9.0  &       4.2 (14.4)&(c)  \\
     A0133  &      4.1  &       1007  & $31.7\pm3.8^d$  &    94.0    &    140    &      100    &      89.0    &      144  &     11.0  &       8.0  &       13.9 (125)&(b)  \\
      A0383  &      4.8  &       944  &     $30.6\pm3.1^d$  &    227    &    108    &      75.5    &      60.8    &      156  &     14.6  &       ...  &       ... &(b)  \\
     A0907  &      6.0  &      1096  & $45.6\pm3.7^d$ &    212    &    167    &      89.8    &      80.2    &      142  &      9.5  &     ...  &       ... &(b)   \\
     A1795  &      6.1  &      1235  &  $60.3\pm5.2^d$&    717    &    106    &      66.8    &      57.5    &      135  &      9.2  &      11.0  &       9.4 (61.7) &(b)   \\
     A1413  &      7.4  &      1299  & $75.7\pm7.6^d$&    244    &    430    &     310    &     301    &      125  &      9.4  &      18.3  &       6.2 (171) &(b)   \\
     A0478  &      7.9  &      1337  & $76.8\pm10.1^d$ &  285    &    336    &     175    &     170    &      152  &      4.7  &       8.4  &       6.6 (209) &(b)   \\
     A2029  &      8.5  &      1362  & $80.1\pm7.4^d$  &  356    &    331    &     230    &     216    &      131  &     13.9  &      20.1  &       7.9 (116) &(b)   \\
     A2390  &      8.9  &      1416  & $107\pm11^d$ &    588    &    842    &     555    &     546    &       87  &      9.5  &       ...  &       ...  &(b)  \\
\end{tabular}
\caption{(1) Designations of the 26 groups and clusters. (2) Mean emission weighted temperature. (3) Radius of last data point. (4) Newtonian dynamical mass at overdensity $\Delta$=1250, 2500, 200 and 500 for the superscripts a, b, c and d respectively. (5) X-ray gas mass at $r_{max}$, may be larger than required for equilibrium. Error is always less than 1\% (6) MOND dynamical mass using the simple $\mu$-function . (7) Residual dark mass required if neutrinos are not present. (8) Contribution of neutrinos to $M_m$ at $r_{max}$. (9) Minimal radius beyond which neutrinos have high enough density to supply remaining dynamical mass. (10) Residual mass if neutrinos contribute maximally. (11) K-band luminosity of BCG. (12) Necessary $M/L_K$ to account for (10). (13) The reference from which the data is taken. (a) indicates Gastaldello et al. (2007a), whereas (b) is Vikhlinin et al. (2006), (c) Zappacosta et al. (2006) and (d) Humphrey et al. (2006). Three clusters overlap between Vikhlinin et al. (2006) and Gastaldello et al. (2007a) which we pair in the table and have similar results.}{18cm}
\end{table*} 

\end{appendix}

\label{lastpage}

\end{document}